\documentclass[12pt]{article}
\usepackage{times}
\usepackage[dvips]{graphicx}

\title{Self-organized wavy infection curve of COVID-19}

\author{
Takashi Odagaki$^{\rm 1,2}$\\
\\
$^{\rm 1}$\normalsize{Kyushu University,}
\normalsize{Nishiku, Fukuoka 819-0395, Japan}\\
$^{\rm 2}$\normalsize{Research Institute for Science Education, Inc.}\\
\normalsize{Kitaku, Kyoto 603-8346, Japan}\\
\normalsize{Corresponding author: t.odagaki@kb4.so-net.ne.jp}
}

\date{\today}

\begin{document} 

\maketitle 

\begin{abstract}
Exploiting the SIQR model for COVID-19, I show that the wavy infection curve in Japan 
is the result of fluctuation of policy on isolation measure imposed by the government and obeyed by
citizens.
Assuming the infection coefficient be a two-valued function of the number of daily confirmed new cases,
I show that when the removal rate of infected individuals is between these two values,
the wavy infection curve is self-organized.
On the basis of the infection curve, I classify the outbreak of COVID-19
into five types and show that these differences can be related to the relative
magnitude of the transmission coefficient and the quarantine rate of infected individuals.

\end{abstract}

\section{Introduction}

Since November 2019, the pandemic COVID-19 is still expanding in the world.
The time dependence of the number of daily confirmed new cases, which I call an infection curve for simplicity,
shows clearly a wavy structure in some countries like USA, Japan, Luxembourg and Sweden \cite{JHU}. 
Since the period of the wave is much shorter than that of the wave observed in the Spanish flu
in 1918-1920, which is believed to be the result of virus mutated while travelling around the globe,
there must be a different origin of the wavy infection curve of COVID-19.
Besides the wavy infection curve, there are several different types of the
infection curve observed in each country in the past 8 months.

Epidemic oscillations have been investigated on the basis of compartmental models \cite{hethcote, brauer}.
Most of approaches attributed the oscillation to a  sinusoidal time dependence of parameters of the model.
Oscillations in SIR models have also been discussed by evolving networks \cite{zhang} and asynchronous
probabilistic cellular automaton \cite{chavez}.
Recently, Greer et al \cite{greer} proposed a simple dynamical model with timevarying
births and deaths to explain sustained periodicity of epidemics like smallpox.
These approaches may not be appropriate to COVID-19 since (1) there are strong effects of measures including
social-distancing on the outbreak, (2) presymptomatic and asymptomatic patients are infectious and
(3) these patients can be identified and quarantined by PCR (Polymerase Chain Reaction) test.

The transmission coefficient of the virus from an infected individual to a susceptible individual
depends on the trait of the virus as well as the frequency of social contact.
Depending on the infection status,  a government imposes a strong measure or a moderate request of lockdown
and citizens reduce social contact among them voluntarily, which introduces fluctuation of
the transmission coefficient.

In this paper, I investigate various infection curves on the basis of SIQR model,
focusing on the fluctuation of the transmission coefficient which depends on the infection status.
In Sec. 2, I first analyze the wavy infection curve of COVID-19 in Japan
and show that it can well be fitted by a transmission coefficient depending on the infection status.
In Sec. 3, I investigate a model society where the transmission coefficient takes two-values
depending on the phase of the outbreak and show that if the sum of quarantine and recovery rates
is between these two values, a wavy infection curve is self-organized.
Analysis of various infection curves in apparent steady states is presented in Sec. 4 where I
classify them into five types which is related to the relative strength of 
lockdown and quarantine measures.
Results are discussed in Sec. 5.

\section{Wavy infection curve in Japan}
\subsection{Model}
In the SIQR model\cite{SIQR, odagaki-idm, suda},
population are separated into four compartments; susceptible individuals, 
infected individuals at large (will be called infecteds for simplicity), 
quarantined patients in hospitals or at home
who are no longer infectious in the community 
and recovered (and died) patients.
The population in each compartment are denoted by $S$, $I$, $Q$ and $R$, respectively,
and the total population is
given by $N (=S+I+Q+R)$.
The basic equations for the time evolution of the populations are
given by a set of ordinary differential equations.
\begin{eqnarray}
\frac{d S}{d t} &=& - \beta S \frac{I}{N}, \\
\frac{d I}{d t} &=& \beta S \frac{I}{N} - q I - \gamma I \\
\frac{d Q}{d t} &=& q I - \gamma' Q ,\\
\frac{d R}{d t} &=& \gamma I + \gamma' Q,
\end{eqnarray}
where $t$ is the time.
The term $\beta S \frac{I}{N}$ denotes the net rate at which infections spread,
where $\beta$ is a transmission coefficient determined by lockdown measure including social-distancing
and self-isolation of people and by the trait of virus.
Infected individuals at large, regardless of whether they are symptomatic or asymptomatic,
are quarantined at a per capita rate $q$ and become non-infectious to the population.
The quarantine rate is determined by the government policy on PCR test.
Quarantined patients recover at a per capita rate $\gamma '$ (where $1/\gamma '$ is the
average time it takes for recovery) and infected individuals at large
become non-infectious at a per capita rate $\gamma$ 
(where $1/\gamma$ is the average time that an infected patient at large
is capable of infecting others).
It is apparent that Eqs. (1) $\sim$ (4) guarantee the conservation of population
$N = S + I + Q + R$.

At the end of November 2020, the total number of infected, quarantined and recovered people
is much smaller than the entire population in any countries, and thus
the pandemic can be regarded as in its early stage far from the stage of herd immunization.
Therefore, I can assume that $I+Q+R << N$ is satisfied and $S= N -(I+Q+R) \simeq N$. 
Then the basic equation governing the time evolution of the number of infecteds
is written as 
\begin{equation}
\frac{d I}{d t} = \beta I - q I - \gamma I \equiv \lambda_I I,
\end{equation}
where the net rate of change of the number of infecteds is denoted as
\begin{equation}
\lambda_I = \beta  - q - \gamma
\end{equation}
which determines the short-term behavior of the number of infecteds.
The number of infecteds increases when $\lambda_I >0$ and decreases when $\lambda_I <0$. 

It is straightforward to obtain the solution to Eq. (5):
\begin{equation}
I(t) = I(0) \exp (\int_0^t \lambda_I(t') dt'),
\end{equation}
where $I(0)$ is the initial number of infecteds.

\subsection{Analysis of the infection curve in Japan}
The observed data for the outbreak of COVID-19 is the daily confirmed
new cases $\Delta Q(t) \equiv qI$, which is given by a convolution of the incubation
period distribution function $\psi(t)$ and the number of infecteds $I(t)$. 
Therefore $\Delta Q(t)$ can be expressed as
\begin{equation}
\Delta Q(t) \propto \int_{-\infty}^t \psi(t-t') I(t') dt'.
\end{equation}
Since the incubation period distribution is a well behaved function with a single
peak \cite{incubperiod,incubationdistri}, 
the convolution can be evaluated by the saddle-point method of integration and 
it is given by\cite{odagaki-idm} 
\begin{equation}
\Delta Q(t) \propto \sqrt{\frac{2\pi}{|\psi''(\tau)|}}\psi(\tau)^{3/2} I(t-\tau),
\end{equation}
where $\tau$ is a characteristic time representing the peak position of $\psi(t)$
and $\psi''(t) = \frac{d^2 \psi(t)}{dt^2}$.
Therefore, defining  $\lambda(t) \equiv \lambda_I(t - \tau)$, I assume that the infection curve
can be written as 
\begin{equation}
\Delta Q(t) = \Delta Q(t_0) \exp (\int_{t_0}^t \lambda(t') dt'),
\end{equation}
and redefine $\beta(t)$ and $q(t)$ so that $\lambda(t)$ can be written
as  $\lambda (t) = \beta(t) - q(t) - \gamma$.

The first wave of the outbreak of COVID-19 in various countries has been analyzed
on the basis of Eq. (10), where $\Delta Q(t)$ is approximated by a piece-wise simple
exponential function \cite{odagaki-idm, suda, italy, india1, india2, sweden, brazil}.

In order to fit the infection curve in Japan by Eq. (10), 
I first assume that $\gamma$ is a constant since no treatment could be given to
infecteds and set $\gamma = 0.04$ \cite{incubperiod,incubationdistri,infectperiod}.

Next, I assume that $\beta(t)$ and $q(t)$ change in time continuously between two values
represented by a hyperbolic tangent function
\begin{equation}
F(x) = A_{if} \tanh\left( \frac{x - x_m}{dx_m} \right) + B_{if} ,
\label{Fofx}
\end{equation}
which satisfies $F(x_i)=F_i$ and $F(x_f)=F_f$ and $x_i \le x_m \le x_f$, namely
\begin{eqnarray}
A_{if} &=&  \frac{F_i - F_f}{\tanh\left( \frac{x_i - x_m}{dx_m} \right) -
          \tanh\left( \frac{x_f - x_m}{dx_m} \right)} ,\\
B_{if} &=& F_i - A_{if} .
\label{AandB}
\end{eqnarray}
Function $F(x)$ changes from $F(x_i) = F_i$ to $F(x_f) = F_f$ continuously
between $x \sim x_m - dx_m$ and $x \sim x_m + dx_m$.

Figure 1 shows the daily confirmed new cases in Japan from March 26 (day 0) to 
November 20(day 239), 2020.
The solid curve in Fig. 1 represents a fitting by piece-wise hyperbolic tangent functions 
for time-dependent transmission coefficient and quarantine rate with fixed $\gamma = 0.04$.

\begin{figure}
\begin{center}
\includegraphics[width=9cm]{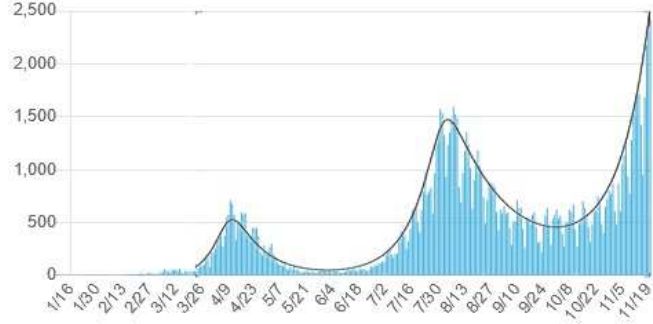}
\caption{The daily confirmed new cases $\Delta Q(t)$ in Japan from March 26 to November 20, 2020.
The solid curve is a fitting by the piece-wise hyperbolic tangent functions for
the transmission coefficient with parameters listed in Table 1.
}
\end{center}
\label{figure1}
\end{figure}

\begin{table}[htb]
\caption{
Parameters for the fitting in Fig. 1. The quarantine rate is increased from q(0) = 0.02 to q(200)= 0.029
as explained in the text and $\gamma = 0.04$ is fixed.
}
\begin{center}
  \begin{tabular}{|c|c|c|c|c|c|c|c|c|c|}
\hline
       & $t_0$ & $t_1$ & $t_2$ & $t_3$ & $t_4$ & $t_5$ & $t_6$ & $t_7$ & $t_8$ \\
\hline
date & 3/26    & 4/12    &  4/30   &  6/04  &  7/19  &  8/03  & 8/23   & 10/12   & 11/21   \\
\hline
days   & 0     & 17    &  35   &  70  &  115  &  130  & 150   & 200   & 240   \\
\hline 
$\beta_{2i}$  & 0.18   &  $-$  & 0.007  &  $-$  & 0.138 &  $-$  & 0.04&  $-$   & 0.127  \\
\hline
$dt_{2i+1}$ &   $-$  & 6     & $-$   &  20  &  $-$  &    6  &  $-$  &  30   &  $-$  \\
\hline
\end{tabular}
\end{center}
\end{table}

The fitting procedure is as follows.
First, I assumed that the quarantine rate satisfies $q (0) = 0.02$ and $q( 200) = 0.029$, and transition occurs at
$t_q = 10$ with width $dt_q = 30$ as given by Eqs. (\ref{Fofx}) $\sim$ (\ref{AandB}).
Then, the transmission coefficient is assumed to be given by Eq. (\ref{Fofx}) for $t_{2i} \le t \le t_{2i+2}$
with $\beta(t_{2i}) = \beta_{2i}$ and $\beta(t_{2i+2}) = \beta_{2i+2}$ and $t_{2i+1}$ as the transition point
and $dt_{2i+1}$ as the width of the transition ($i = 0, 1, 2, \dots$).
Table 1 summarizes parameters determining the time dependence of $\beta(t)$ used for fitting in Fig. 1. 

Figure 2(a) shows the time dependence of $\beta(t)$, $q(t)$ and $\lambda(t)$
and Fig. 2(b) shows a parametric plot of $\beta(t)$ as a function of $\Delta Q(t)$.
\begin{figure}
\begin{center}
\includegraphics[width=6cm]{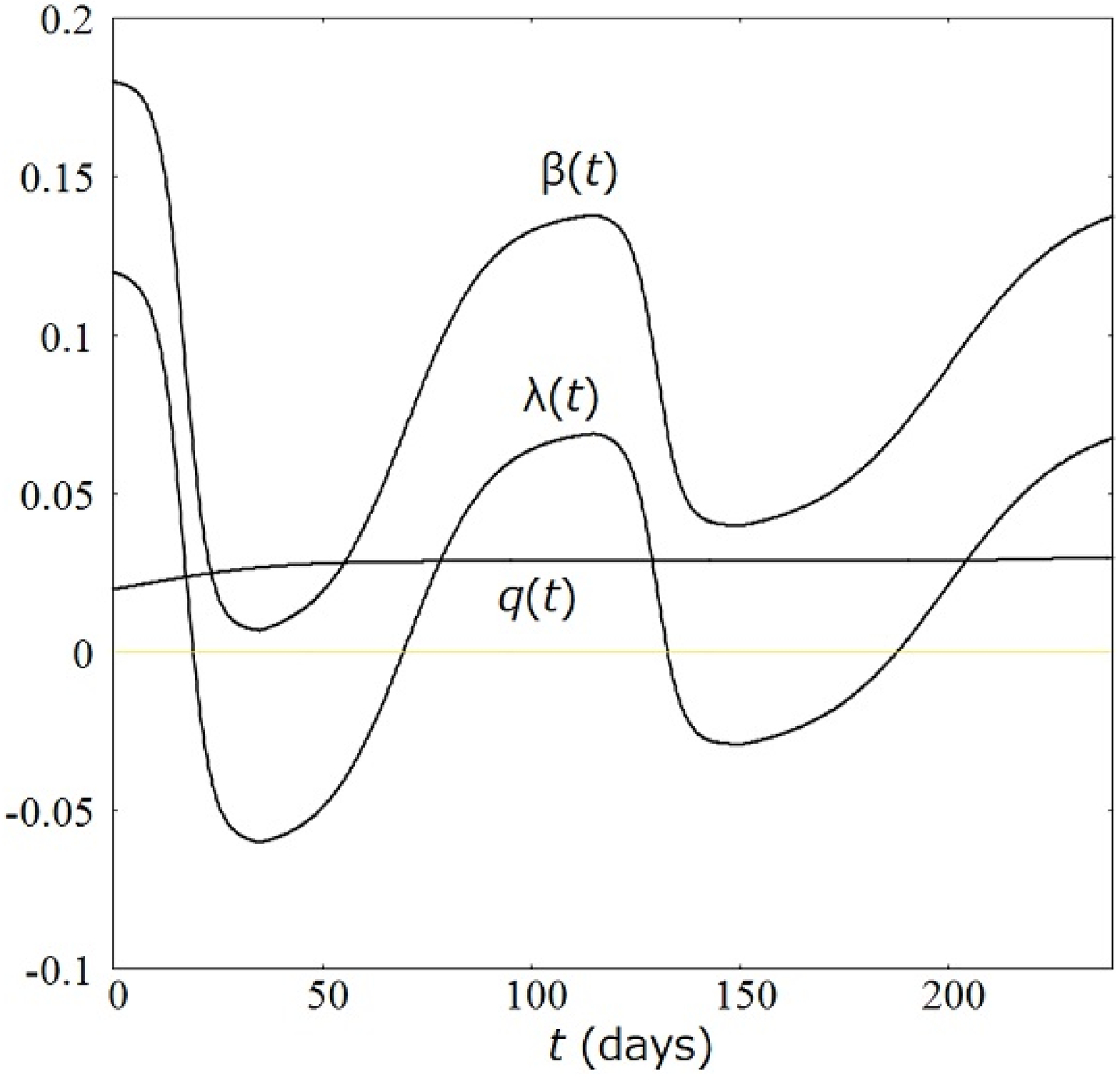} \hspace{1cm}
\includegraphics[width=6.3cm]{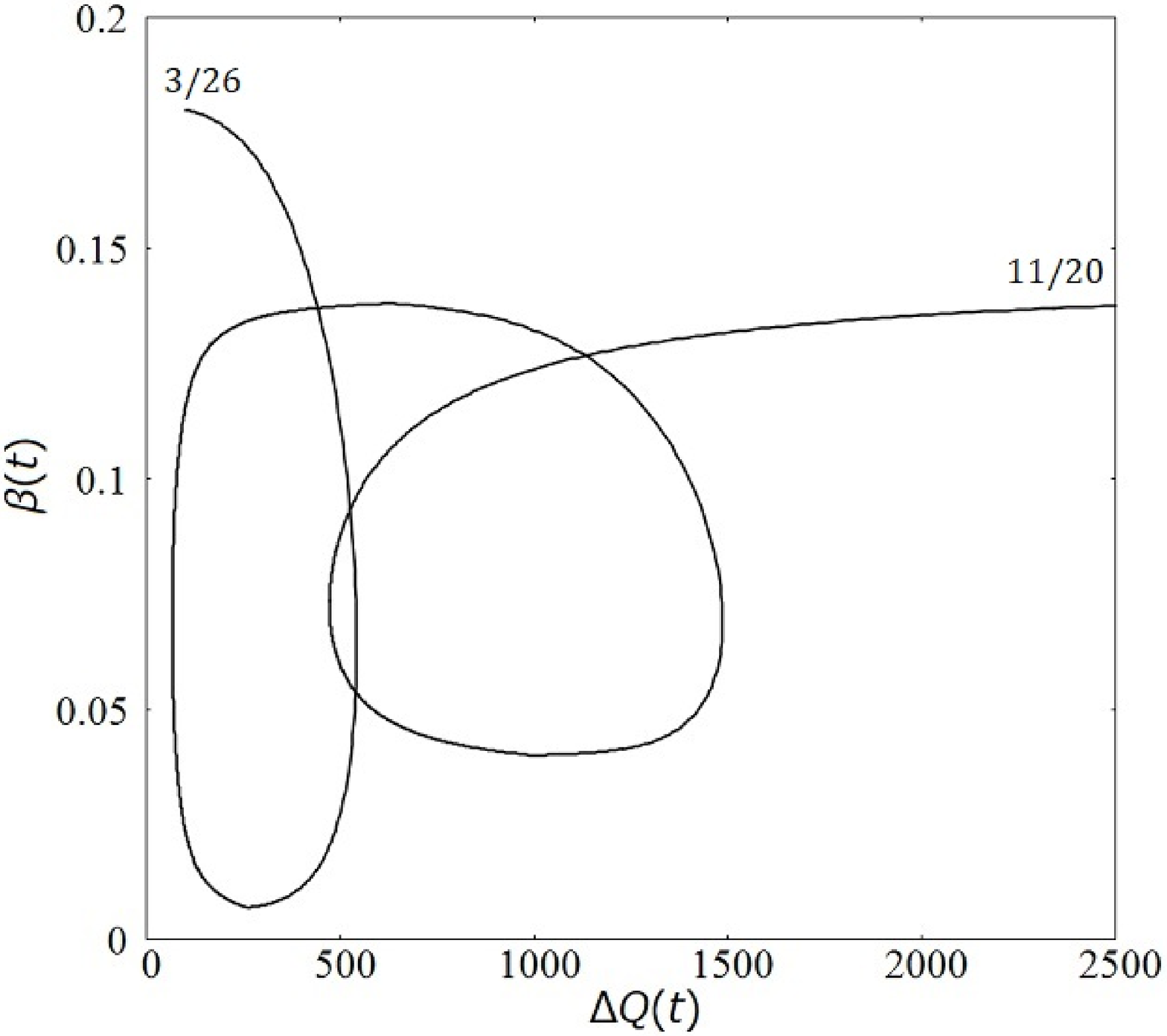} \\
(a) \hspace{6.5cm}(b)
\caption{(a) Parameters for the transmission coefficient $\beta(t)$.
$\beta(t) = A \tanh\left( \frac{t - t_{2i+1}}{dt_{2i+1}}\right) + B$ is set to satisfy
$\beta(t_{2i}) = \beta_{2i}$ and $\beta(t_{2i+2})= \beta_{2i+2}$ for $t_i$'s listed in Table 1.
(b) Parametric plot of $\beta(t)$ as a function of $\Delta Q(t)$ for Japan from March 26 to November 20, 2020.
}
\end{center}
\label{figure2}
\end{figure}
It should be emphasized that the assignment of $\beta$ and $q$ from $\lambda$ is not unique since 
adding any amount to $\beta$ and $q$ at a given time does not change $\lambda$.
In the present study, I assumed that the time dependence of $q$ is weak since the procedure for the PCR
test did show no drastic change in the period for identifying infected individuals at large.

\section{Self-organization of wavy infection curve}
The time dependence of the transmission coefficient $\beta(t)$
must be attributed to the attitude of people to self-isolation under government policy and massive
information from news media.
As Fig. 2(b) indicates,  $\beta(t)$ depends strongly on $\Delta Q(t)$.
Therefore, I consider the transmission coefficient be a function of $\Delta Q$ and $\frac{d \Delta Q}{dt}$,
and I introduce a model country in which $\beta$ is given by
\begin{equation}
\beta (\Delta Q) = \left\{ \begin{array}{ll}
\beta_h &\quad  \mbox{when $\frac{d\Delta Q}{dt} > 0$ and $\Delta Q \le \Delta Q_h$ } \\
\beta_{\ell} &\quad  \mbox{when $\frac{d\Delta Q}{dt} < 0$ and $\Delta Q_{\ell} \le \Delta Q < \Delta Q_h$ } ,\\
\end{array}
\right.
\label{betaofi}
\end{equation}
where $\beta_{\ell} < q + \gamma < \beta_h$ is satisfied. Figure 3 (a) shows $\beta (\Delta Q)$.

The infection curve for $\beta (\Delta Q)$ given by Eq. (\ref{betaofi}) is shown in Fig. 3 (b), where I set $\Delta Q(0) = \Delta Q_0 = 10$,
$\Delta Q_h = 100  $ and $\Delta Q_{\ell} = 20$, and $\beta_h = 0.15 $, $\beta_{\ell} = 0.05  $ and $q = 0.05  $, $\gamma = 0.04$.
In this plot, $\Delta Q_{\ell}$ is set to  0 after the third wave. 
The infection curve clearly shows wavy nature. If $\Delta Q_{\ell}$ is kept at the same value after the third wave, the wavy
infection curve continues.
\begin{figure}
\begin{center}
\includegraphics[width=5.5cm]{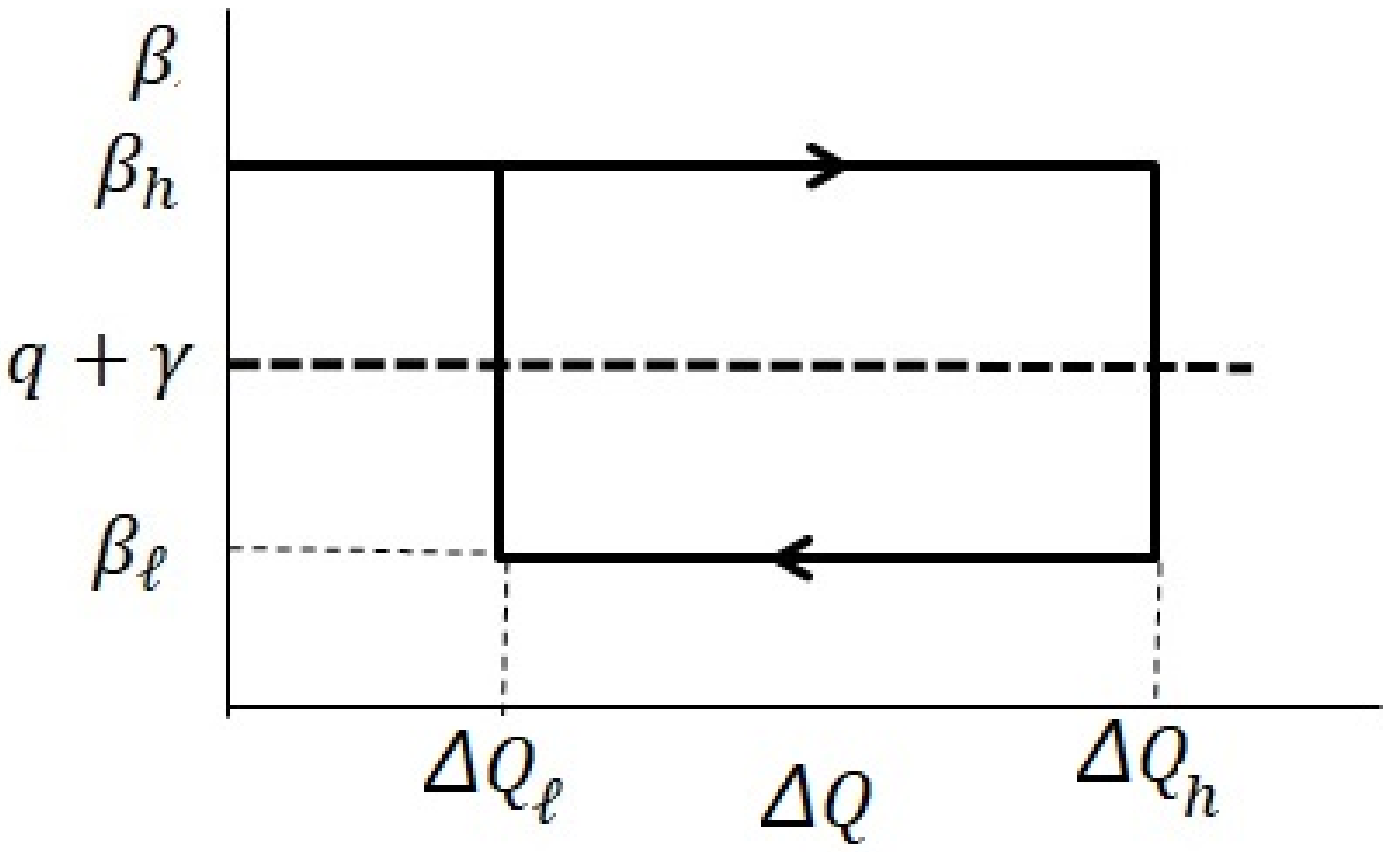}\hspace{1cm}
\includegraphics[width=5.5cm]{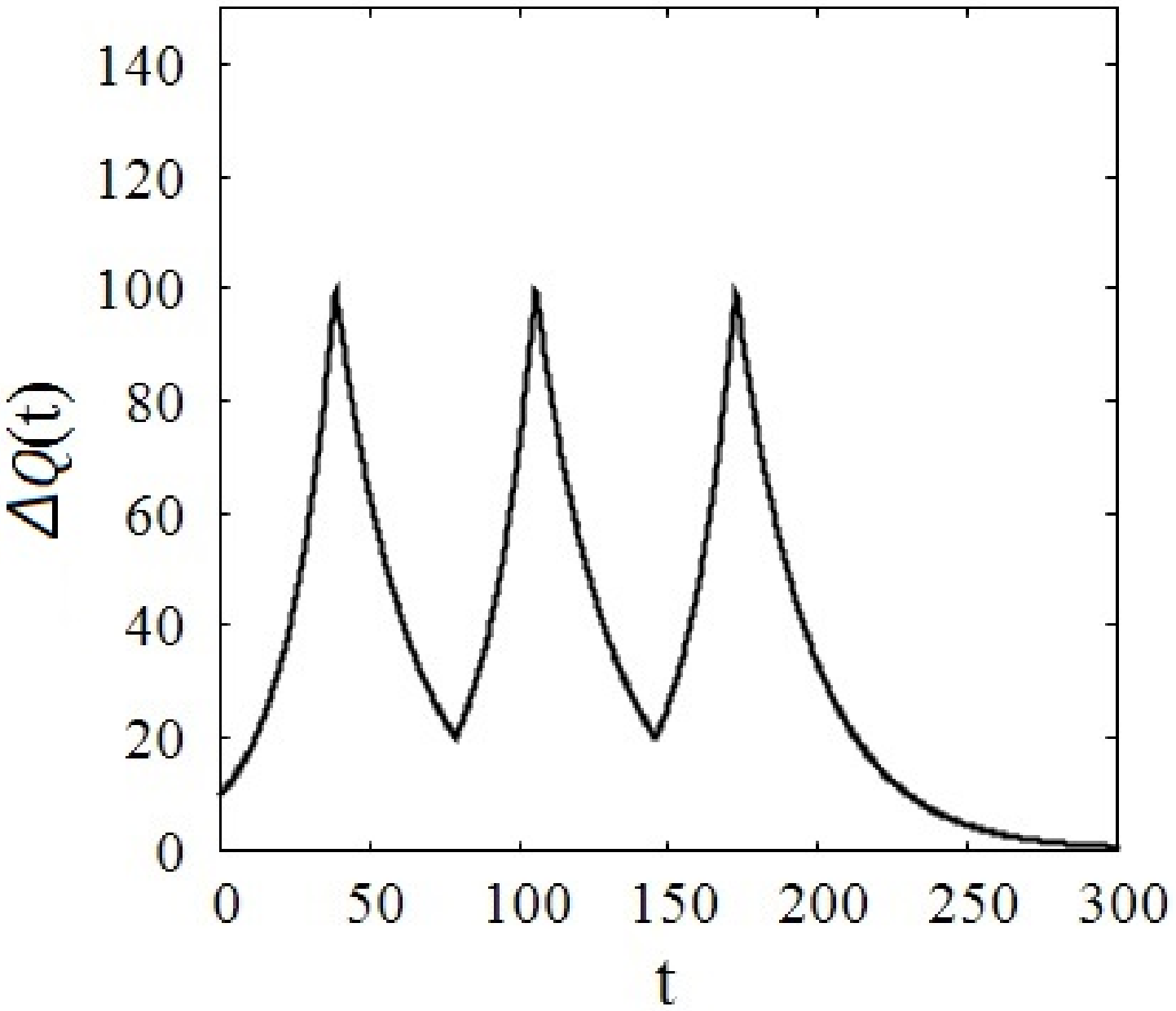}\\
\hspace*{0.5cm}(a)\hspace{6.5cm}(b)
\caption{(a) Model transmission coefficient depending on $\Delta Q$.
(b) Wavy infection curve for the model transmission coefficient.
 }
\end{center}
\label{figure3}
\end{figure}
If $\beta(\Delta Q)$ is given by a continuous function, then the infection curve becomes a smooth function
as in Fig. 1.

\section{Assessment of measures }

According to the data available at Coronavirus Resource Center, Johns Hopkins University \cite{JHU},
the steady behavior of the infection curve in each country up to November 19th
seems to be classified into the following five types:
\begin{description}
\item [Type I] The infection curve keeps increasing, like Jordan, Ukraine and Morocco.
\item [Type II] After some number of peaks, the infection curve increases again like Type I.
This is seen in Japan,  USA, Russia, Canada and many European countries. This can be changed to Type III by some measures.
\item [Type III] The infection curve shows oscillation like in UAE, Finland and Ireland. 
Countries in this type move easily to Type II unless strong measures are introduced to move to Type V.
\item [Type IV] Infection curve is characterized by a sharp peak followed by more or less constant
infection for a long time. This infection curve is seen in Equador, Kuwait and Honduras.
Countries in this type usually move to either Type II or Type III, though they could move to Type V.
\item [Type V] After a small peak, few new cases are observed like in China, Taiwan, Thailand
and Viet Nam.
\end{description}

As discussed in Sec. 3, the relative magnitude of $\beta$ and $q+\gamma$ must be responsible
for the structure of the infection curve. Here, keeping $q+\gamma$ constant, I discuss
the relative magnitude of these parameters for different types of infection curve.
It should be emphasized that the difference $\beta - (q+\gamma)$ determines the infection curve.
For the sake of simplicity, I fix $q+\gamma$ and attribute all effects to change in $\beta$.
It is possible to discuss in the same way by changing $q$ with a fixed $\beta$.

For Type I infection curve,  $\beta > q + \gamma$ is satisfied (Fig. 4(a))
and thus the infection curve keeps increasing.
The infection curve will reach eventually its maximum and start to decline because of
the non-linear term $SI/N$ in Eqs. (1) and (2). The Spanish flu belongs to this type.

Type II infection curve will be realized when a strong lockdown measure is introduced at the outbreak
and it is lifted in fear of economic break down (Fig. 4(b)). 
After a little peak and some length of tail, the infection curve will follow the same trend as Type I.

Wavy infection curve (Type III, Fig. 4 (c)) has already been discussed in Sec. 3.

Infection curve of Type IV is characterized by a fixed point in the $\beta$ - $\Delta Q$ plane which is reached
after the first peak (Fig. 4(d)), and $\Delta Q_{\ell}$ determines the size of the daily confirmed new cases.

Type V infection curve represents the most efficient measure; the transmission coefficient
is brought below $q+\gamma$ (or $q$ is increased) so that $\beta < q + \gamma$ is satisfied,
and the measure is kept. The trajectory in the $\beta$ - $\Delta Q$ plane has a fixed point
near $\Delta Q = 0$ as shown in Fig. 4(e).

\begin{figure}
\begin{center}
\includegraphics[width=4cm]{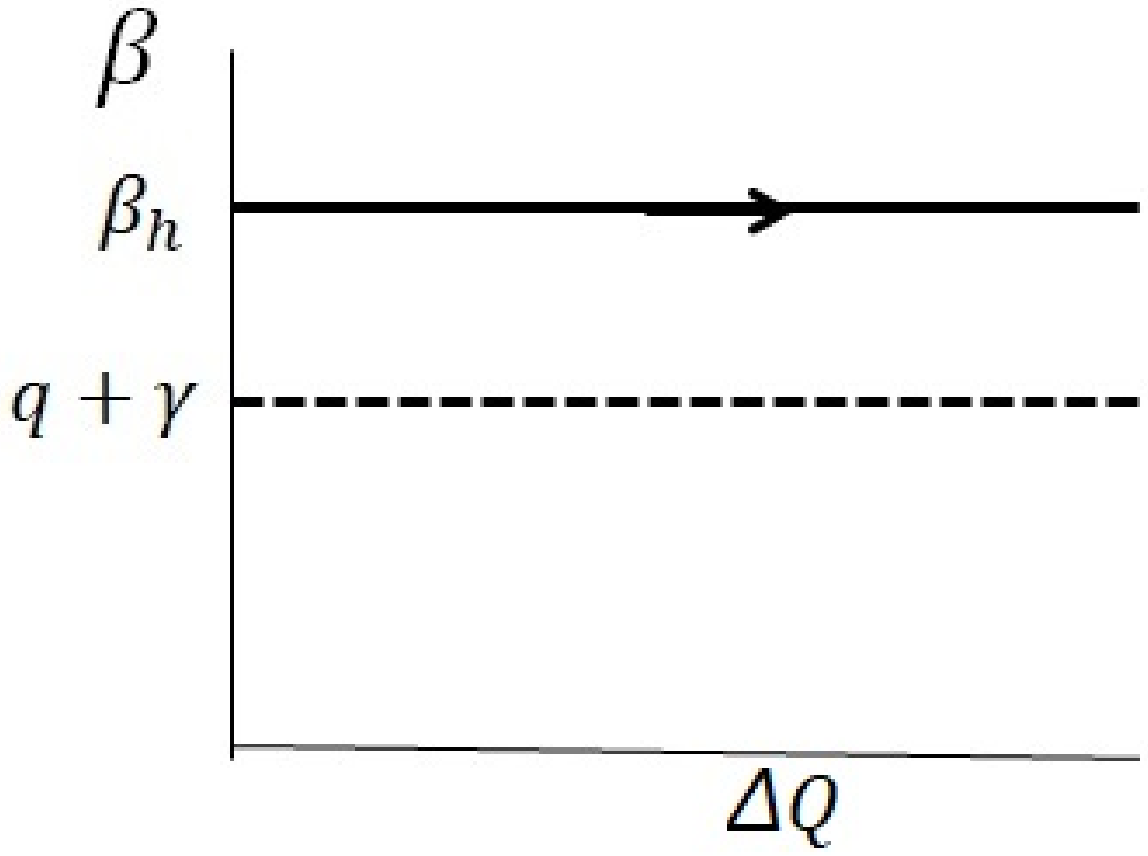}\hspace{1cm}
\includegraphics[width=4cm]{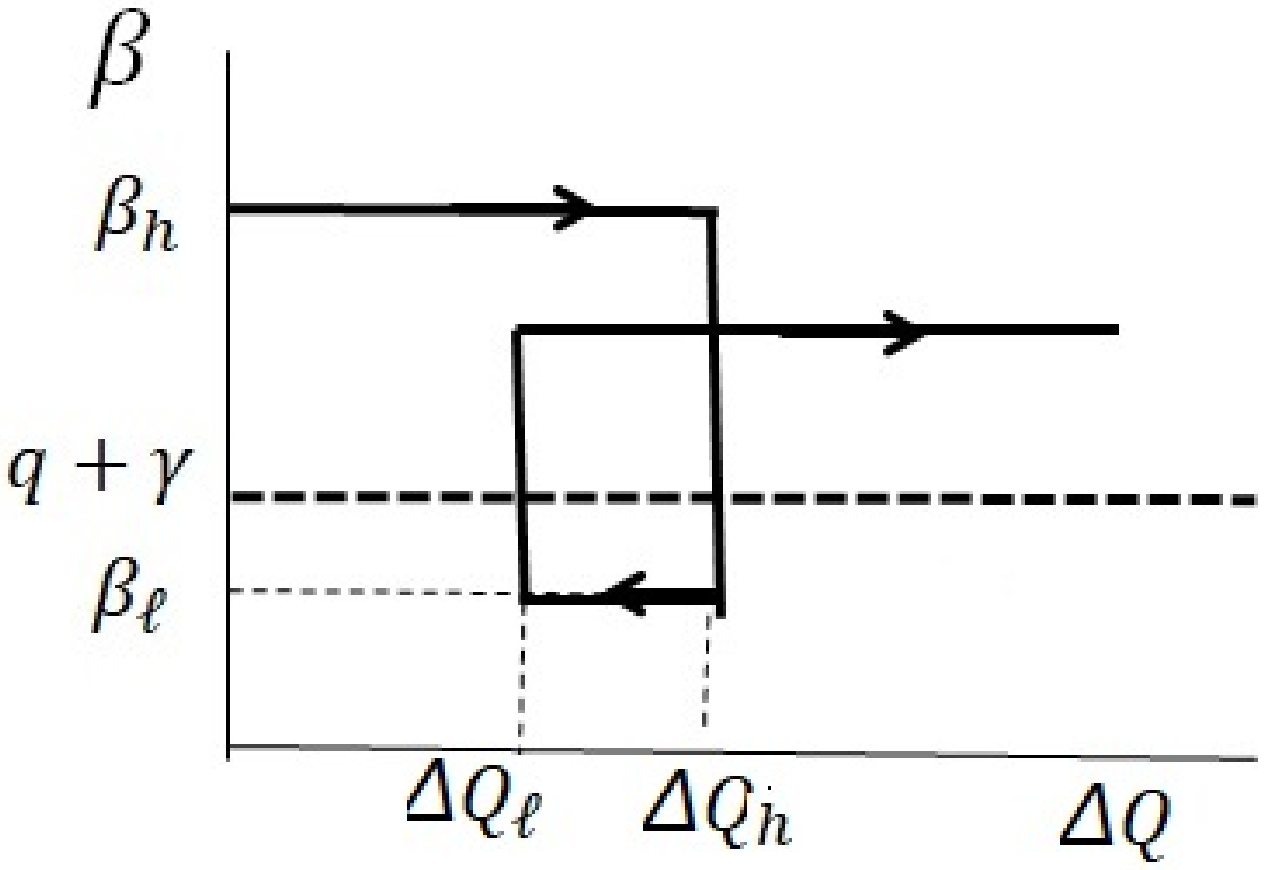}\\
\hspace*{0.5cm}(a)\hspace{4cm}(b)\\
\includegraphics[width=4cm]{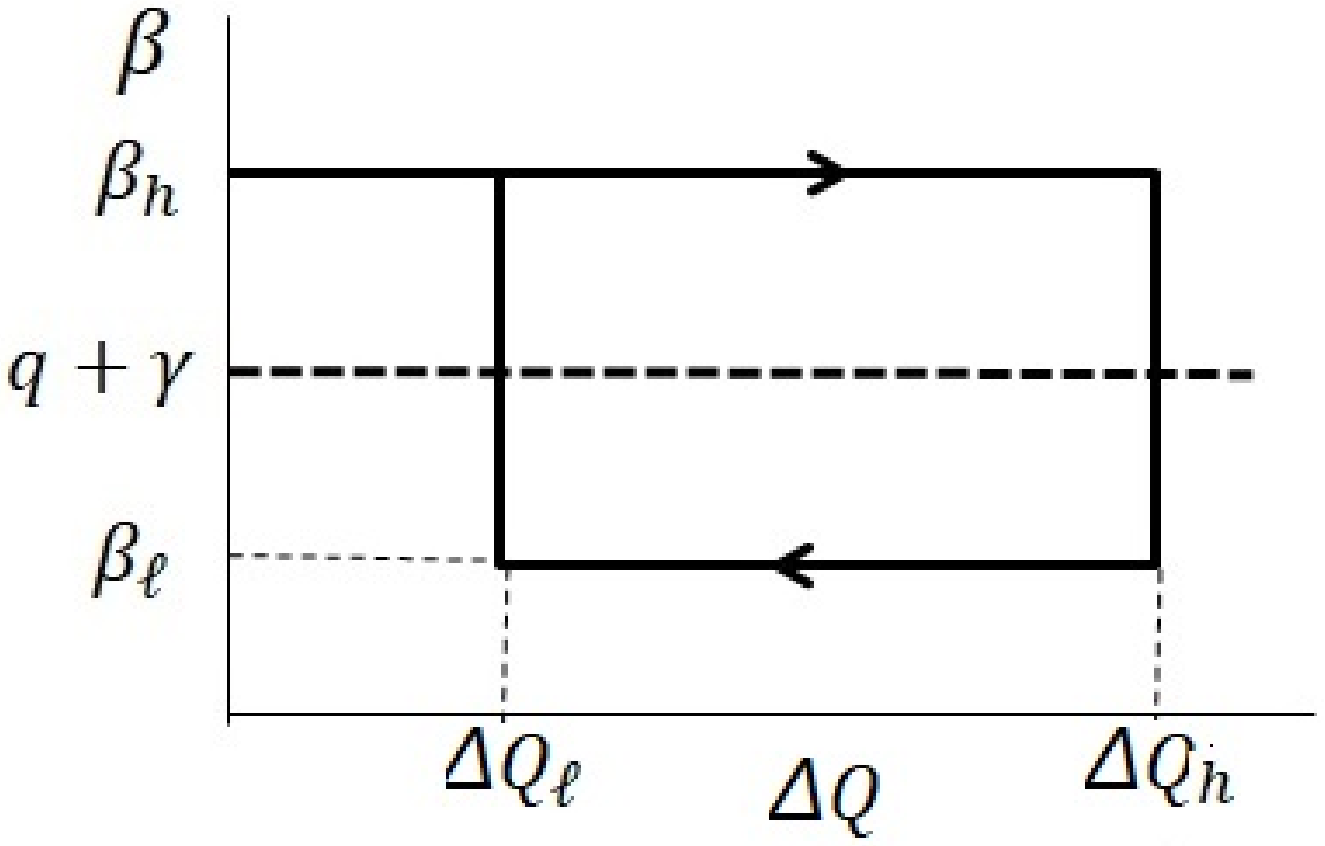}\hspace{1cm}
\includegraphics[width=4cm]{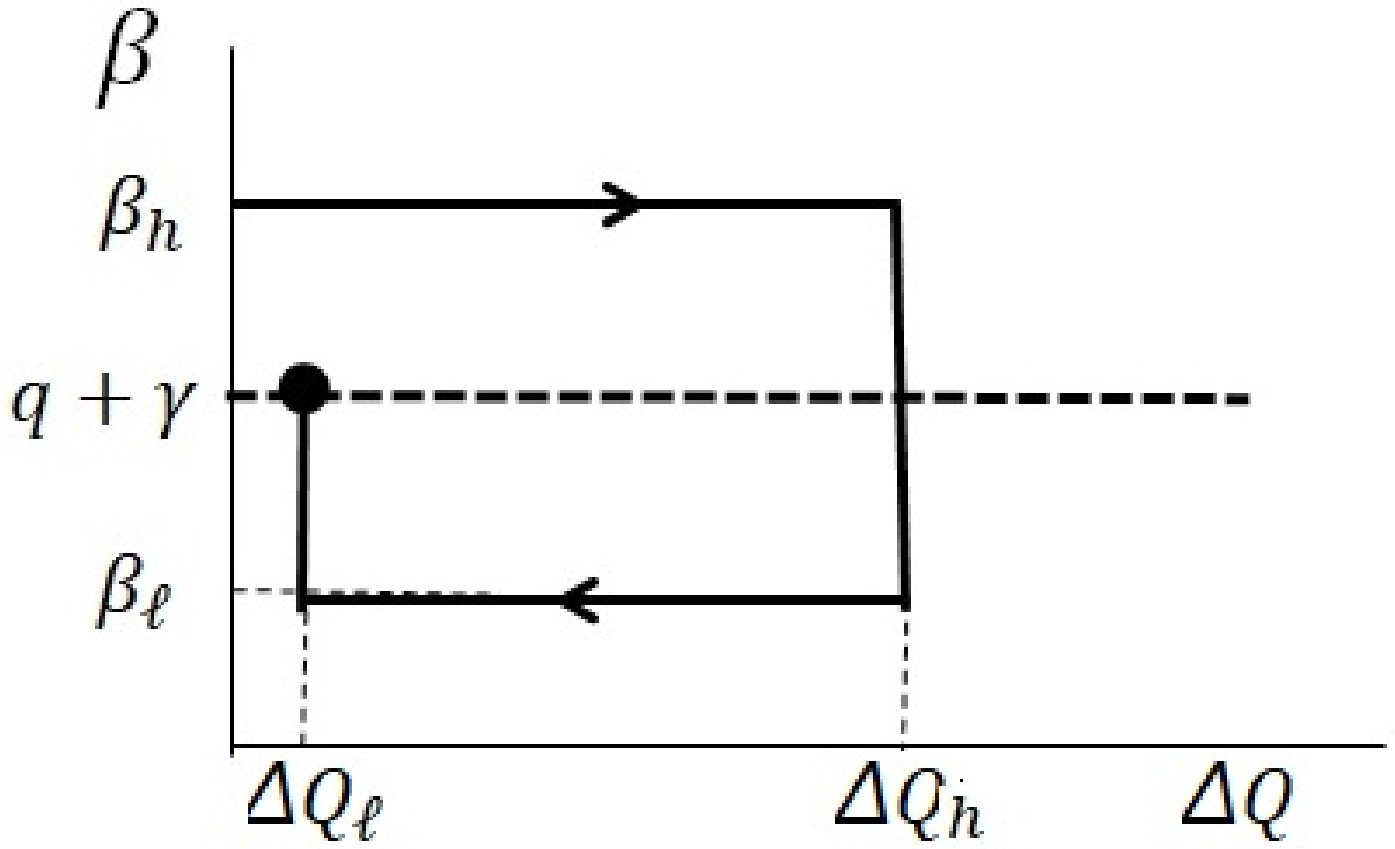}\hspace{1cm}
\includegraphics[width=2.5cm]{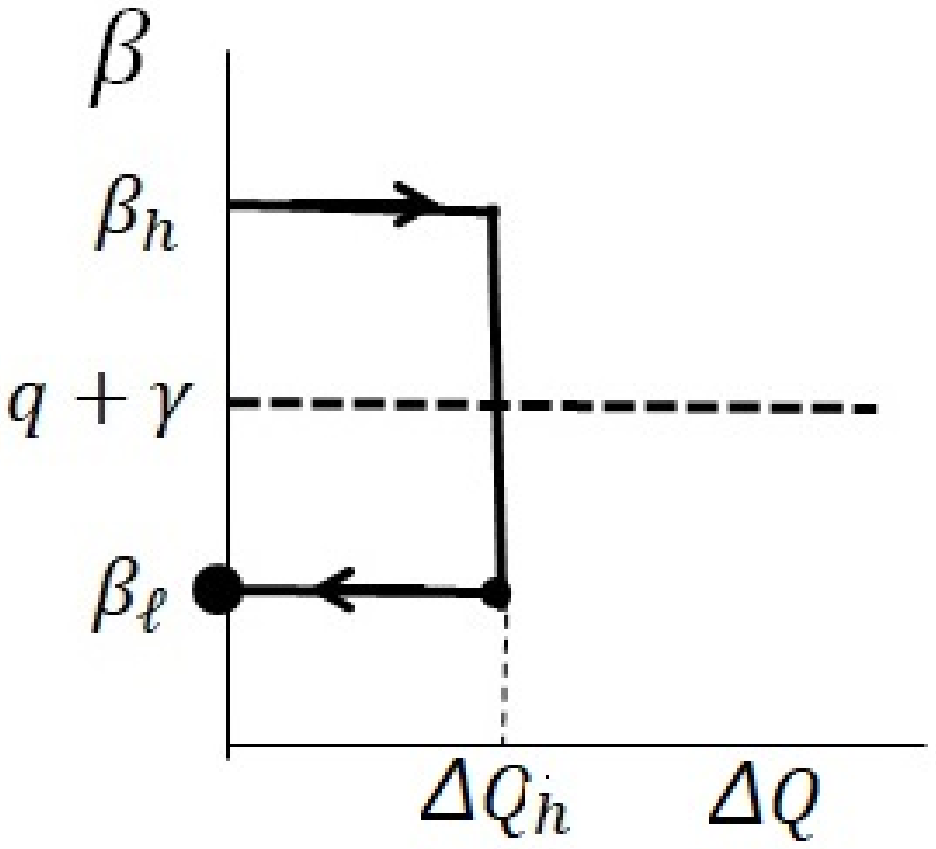}\\
\hspace*{0.5cm}(c)\hspace{4cm}(d)\hspace{4cm}(e)
\caption{Expected relative magnitude of $\beta(\Delta Q)$ and $q+\gamma$, which is
shown for fixed $q+\gamma$.
 (a) Type I, (b) Type II, (c) Type III, (d) Type IV and (e) Type V.
}
\end{center}
\label{figure4}
\end{figure}

\section{Discussion}
In this paper, I discussed the infection curves of COVID-19 observed in many countries
and showed that the infection curve in an apparent steady state can be classified into five types.
In particular, a wavy infection curve can be self-organized due to change in self-isolation
and/or quarantine measures making $\beta$ above or below $q+\gamma$.
It is shown that these different infection curves are caused by relative strength
of lockdown measure and quarantine measure.
It should be emphasized that the infection curve is determined by the interplay between tranmission of the virus
and quarantine of patients, and thus unless loosening of lockdown measures is compensated by
strengthening of quarantine measures, the infection will continue to expand.

It will be possible to formulate the optimum policy specific to the country for controlling the outbreak
on the basis of the present theoretical framework, if the cost function and the aim of policy in each country
are given \cite{SIQRexact}.

The pandemic in countries whose infection curve is of Type I or Type II 
will stamp out when sufficient number of population get immunized.
According to percolation theory \cite{Odagakibook}, the condition for the herd immunity is that
the fraction of immunized individuals is larger than a critical value
\begin{equation}
p_c = 1 - \frac{4.5}{n\beta},
\end{equation}
where $\beta$ is the transmission coefficient and $n$ is the average number of
people with whom an infected individual meets while it is infectious.
The critical value depends on $\beta$ and $n$ and it could be as large as $50 \sim 80$ \%.
Therefore, it could take much longer time before the herd immunity
is realized in any countries in the world.


\end{document}